\documentclass[%
 aip,
 amsmath,amssymb,
 reprint,%
]{revtex4-1}

\usepackage{graphicx}
\usepackage{dcolumn}
\usepackage{bm}
\usepackage{color}
\usepackage{comment}
\usepackage{url}
\usepackage[normalem]{ulem}
\usepackage[version=3]{mhchem} 

\usepackage{amsmath}
\usepackage{amssymb}
\usepackage{lipsum}
\usepackage{url}
\usepackage[breaklinks]{hyperref}
\usepackage{comment}
\newcommand{\figurewidth}{.45\textwidth}
\renewcommand{\epsilon}{\varepsilon}
\graphicspath{{dpd-figs/}}

\newcommand{\tn}[1]{\textnormal{#1}}  

\newcommand*{\vv}{\mathbf v}
\newcommand*{\rr}{\mathbf r}
\newcommand*{\hrr}{\hat{\mathbf r}}

\begin{document}

\title{Dissipative particle dynamics for coarse-grained models}

\author{Tine Curk}
\email{tcurk@jhu.edu}
\affiliation{Department of Materials Science \& Engineering, Johns Hopkins University, Baltimore, Maryland 21218, USA}


\begin{abstract}
We develop a computational method based on Dissipative Particle Dynamics (DPD) that introduces solvent hydrodynamic interactions to coarse-grained models of solutes, such as ions, molecules, or polymers. DPD-solvent (DPDS) is a fully off-lattice method that allows straightforward incorporation of hydrodynamics at desired solvent viscosity, compressibility, and solute diffusivity with any particle-based solute model. Solutes interact with the solvent only through the DPD thermostat, which ensures that the equilibrium properties of the solute system are not affected by the introduction of the DPD solvent, while the thermostat coupling strength sets the desired solute diffusivity. Thus, DPDS can be used as a replacement for traditional molecular dynamics thermostats such as Nos\'e-Hoover and Langevin. We demonstrate the applicability of DPDS in the case of polymer dynamics and electroosmotic flow through a nanopore. The method should be broadly useful as a means to introduce hydrodynamic interactions to existing coarse-grained models of solutes and soft materials.
\end{abstract}

\maketitle

Coarse-grained implicit-solvent models are usually parametrized to produce a desired equilibrium distribution of states. To use the same models in non-equilibrium simulations with desired transport properties and hydrodynamic behavior requires
a method that introduces hydrodynamic interactions at the specific solvent viscosity, compressibility and solute diffusivity without affecting the equilibrium configurational properties of the system. 

To achieve this, the coarse-grained system can be coupled to a mesh, typically a cubic grid, and hydrodynamics is solved using this mesh. A few examples of such approaches include Multi-Particle Collision dynamics (MPC) or Stochastic Rotation Dynamics~\cite{kapral1999,gompper2009}, Lattice Boltzmann (LB) methods~\cite{kruegerLB2016}, and Fluid Particle Dynamics (FPD)~\cite{tanaka2000}.  However, coupling of the continuous system to a mesh can introduce artifacts and increase propagation errors. MPC does not conserve angular momentum (or breaks time-reversal symmetry) and suffers from artificially high compressibility~\cite{gompper2009}. LB allows embedding of large colloids, but embedding of particles smaller than the lattice size, \textit{e.g.}, polymers, suffers from lattice discretization errors and mixing molecular dynamics with LB propagation leads to energy conservation issues~\cite{dunweg2009}.  FPD directly solves the Navier--Stokes equations but can only embed objects at least a few lattice sites in size, which makes the method suitable for nanoparticles and colloids~\cite{akira2018}, but is computationally inefficient for small molecules, ions or polymers.

Among off-lattice approaches the most straightforward way to partially introduce hydrodynamics 
is to apply a Galilean-invariant thermostat directly to the solute particles~\cite{soddemann2003}, however, this approach requires a dense suspension and is not applicable to dilute solutions. 
A general off-lattice method is Dissipative Particle Dynamics (DPD) that simulates hydrodynamic interactions by representing the fluid as soft particles~\cite{hk1992,groot1997}. 
DPD reproduces a desired solvent viscosity and compressibility while describing thermal fluctuations and allowing the explicit representation of different chemical components~\cite{espanol2017}. 
However, simply adding DPD particles to a coarse-grained system introduces solvation and depletion effects, and thus changes the equilibrium properties of the system. To overcome these problems, the solute model must be designed for a specific DPD solvent parametrization and typically the solute--solute interaction is restricted to the soft DPD repulsive interaction, which significantly limits the space of solute models. For example, point charges cannot be included due to electrostatic divergence issues when combined with the DPD soft repulsive potentials~\cite{groot2003}. While it is possible to model a variety of multicomponent systems by modifying the conservative interactions between DPD particles~\cite{pagonabarraga2001,groot2001,Guo2011}, this approach requires careful parametrization; altering conservative interactions does not merely change the equilibrium properties but also transport behavior such as local viscosity, compressibility and diffusivity. 
  
Here we propose an alternative off-lattice method, named DPD solvent (DPDS), that incorporates hydrodynamic interactions in coarse-grained models of solutes.
The premise of the method is to overlay the DPD fluid at desired viscosity and compressibility with the solute system. The DPD solvent interacts with the solute system only \textit{via} the DPD thermostat. Thus, the equilibrium properties of the system and the DPD-solvent are fully decoupled and can be controlled separately,
while the transport properties are determined by the solute--solvent thermostat coupling strength. 
The main advantage of this procedure over existing lattice methods is that it avoids any lattice mapping issues. The same or a similar velocity-Verlet integrator is used to evolve configurations both for the solute system and the DPD solvent, which minimizes propagation errors due to solvent--solute coupling and significantly simplifies the implementation compared to hybrid lattice approaches. The main advantage of DPDS over standard DPD is that DPDS can be directly used with any solute model and is not restricted to soft DPD repulsive potentials. This enables straightforward simulations of charged particles and introduction of hydrodynamic interactions to already developed solute models. Moreover, hydrodynamic interactions can be switched on/off without affecting any equilibrium properties of the solute system, allowing direct evaluation of hydrodynamic effects compared to, for example, Langevin dynamics. 

In the following, we review the DPD technique and then show how to combine it with coarse-grained systems of solutes using the DPDS approach. We demonstrate the applicability of DPDS on two typical coarse-grained models, a bead-spring polymer and an electrolyte solution, and show that it accurately describes hydrodynamic interactions by investigating polymer collapse, Zimm dynamics and electrokinetic flow in patterned nanochannels. 

\section{review of the DPD fluid properties}

The DPD fluid is modeled as a canonical ensemble of soft spherical particles at density~$\rho$ and temperature~$T$. The particles interact \textit{via} three distinct interactions. The conservative pair interaction is typically chosen as parabolic repulsion, with the potential energy of an ensemble of $N$ particles
\begin{equation}
V(\mathbf{r}^N)=\frac{1}{2} \sum_{ij} \begin{cases}
 a_{ij}(1-r_{ij}/r_\tn{c})^2\;, & \tn{for }  r_{ij}<r_\tn{c} \\
0\;, & \text{otherwise} \;.
\end{cases}
\label{eq:V}
\end{equation}
The sum proceeds over all particle pairs $i,j$ with $\mathbf{r}_i$ the position of particle $i$, $r_{ij}=|\mathbf{r}_i-\mathbf{r}_j|$ the inter-particle distance and $r_\tn{c}$ the cutoff length. The prefactor $a_{ij}$ is set to reproduce the desired compressibility of the fluid~\cite{groot1997} and can be different among different particle types to model multicomponent fluids.

The dissipative friction force between particles $i$ and $j$ is
 \begin{eqnarray}
F^\tn{d}_{ij}&=& \gamma \omega(r_{ij})^2(\mathbf{v}_{ij} \cdot \hrr_{ij})\hrr_{ij}  \nonumber \\ &-&  \gamma_\perp \omega_\perp(r_{ij})^2( \mathbf{I} - \hrr_{ij} \hrr_{ij}^T) \mathbf{v}_{ij}   \;.
\label{eq:Fd}
\end{eqnarray}
where the first term describes the friction force parallel to the inter-particle vector $\rr_{ij}=\rr_j-\rr_i$ with $\hrr_{ij}=\rr_{ij}/|\rr_{ij}|$ the corresponding unit vector. The friction is proportional to the coefficient $\gamma$ and distance-dependent weighting function $w(r_{ij})$.  The second term describes the corresponding perpendicular contribution~\cite{junghans2008}, \textit{i.e.}, the shear friction, with coefficient $\gamma_\perp$ and weighting function $w_\perp(r_{ij})$.  $\mathbf{I}$ is the identity matrix and  $\hrr_{ij}^T$ denotes the transpose of $\hrr_{ij}$.

The fluctuation--dissipation relation implies that the random force between the two particles is given by
\begin{eqnarray}
F^\tn{r}_{ij}&=&\sqrt{2\gamma k_{\tn B}T} \omega(r_{ij}) \frac{dW_{ij}}{dt} \hrr_{ij}  \nonumber \\
&+& \sqrt{2\gamma_\perp k_{\tn B}T}  \omega_\perp(r_{ij})  \frac{d\mathbf{W}_{ij}}{dt} (\mathbf{I} - \hrr_{ij} \hrr_{ij}^T)  \;,
\label{eq:Fr}
\end{eqnarray}
where the prefactors are determined by the desired temperature~$T$ of the system, $dW_{ij}$ is the independent increment of a Wiener process and $d{\mathbf W}_{ij}$ the vector of independent Wiener processes. For a finite time-step~$\Delta t$, $\frac{\Delta W_{ij}}{\Delta t}=\zeta_{ij} \Delta t^{-1/2}$, and $\frac{\Delta {\mathbf W}_{ij}}{\Delta t}=\boldsymbol{\zeta}_{ij} \Delta t^{-1/2}$ where $\zeta_{ij}$ is a symmetric random variable with zero mean and unit variance and $\boldsymbol{\zeta}_{ij}$ is the corresponding random vector.

The stochastic differential equations that determine the evolution of the fluid are thus
\begin{eqnarray}
m_i \frac{d\vv_i}{dt} &=& -\frac{\partial V}{\partial \rr_i} + \sum_jF^\tn{d}_{ij} + F^\tn{r}_{ij} \;, \nonumber \\
 \frac{d\rr_i}{dt}&=&\vv_i \;.
\end{eqnarray}
with $m$ the mass of the DPD particles.
These equations are usually evolved using the velocity-Verlet integrator~\cite{groot1997,pagonabarraga1998} and the DPD model is implemented in various open-source molecular dynamics (MD) packages such as LAMMPS, EPResSo and HOOMD-blue. Since the dissipative and random forces act only on relative velocities, the integrator preserves Galilean invariance and conserves momentum, implying that the behavior of the fluid follows Navier-Stokes hydrodynamics on sufficiently large length scales. 

When applied to molecular simulations, the main advantage of DPD compared to Langevin dynamics or finite element hydrodynamic solvers is that the DPD method simultaneously samples the correct canonical $(N,V,T)$ Gibbs-Boltzmann ensemble, preserves hydrodynamics and can capture chemical specificity. Consequently, the method has been widely used to simulate hydrodynamic interactions in various systems including colloidal suspensions, blood, phase-separating fluids, polymer solutions, electrolytes, and biological membranes; see Refs.~\cite{espanol2017,santo2021} for extensive reviews of DPD fundamentals and applications.

In the following we use standard parameters corresponding to an aqueous solution with each DPD particle representing $N_\tn{w}=3$ water molecules: the particle density~$\rho_{\tn s}=3/r_\tn{c}^3$, with the cutoff $r_c=0.646\,$nm, which is taken as the length unit~$\lambda=r_\tn{c}$, and the interaction prefactor $a_{ij}=78k_{\tn B}T$ reproduces the compressibility of water at room temperature~\cite{groot1997}.
To reduce the number of parameters, the dissipative weighing function~$w(r_{ij})$ has the same form as the conservative repulsive force, a common choice for DPD simulations~\cite{espanol2017}, 
\begin{equation}
w(r_{ij}) = 1-\frac{r_{ij}}{r_\tn{c}} \;.
\label{eq:w}
\end{equation}

The friction coefficient~$\gamma$ determines the crossover timescale between the ballistic and diffusive regimes~$\tau_\tn{BD} \sim m/\gamma$ (exact calculation is provided below). The typical choice for the friction parameter is $\gamma=4.5 \sqrt{k_\tn{B}T m}/r_\tn{c}$~\cite{groot1997}, which ensures that the dynamics is diffusive on all relevant length scales ($r \ge r_\tn{c}$) and time scales ($t \ge \tau$), where the molecular dynamics time unit~$\tau$, which measures the characteristic time required to move ballistically by a distance $r_\tn{c}$, is $\tau=r_\tn{c} \sqrt{m/(k_\tn{B}T)}$. The original DPD formulation used no shear friction ($\gamma_\perp=0$)~\cite{groot1997}, but its inclusion has been shown to improve hydrodynamic properties~\cite{junghans2008} so we use $\gamma_\perp=\gamma$ and $w_\perp(r_{ij})=w(r_{ij})$~\cite{lauriello2021}.

We note that employing these standard DPD parameters contains an inherent limitation that the Schmidt number $\tn{Sc}=\nu_\tn{k}/D$, which measures the ratio of momentum diffusion to mass diffusion, with $\nu_\tn{k}$ the kinematic viscosity and $D$ the diffusion constant, is too low. $\tn{Sc}\sim 7$ for the DPD fluid, which is much smaller than $\tn{Sc}~\sim 370$ expected for water at room temperature ($T=298\,$K) and standard pressure~\cite{groot1997,lauriello2021}. This discrepancy is a consequence of using a soft repulsive potential [Eq.~\eqref{eq:V}] and relatively weak friction that together allow using significantly larger time-steps, which increases the efficiency of the method by about four orders of magnitude compared to atomistic simulations~\cite{groot2001}. 
However, the low Schmidt number does not appear to affect transport properties of polymers as long as momentum diffusion is faster than mass diffusion ($\tn{Sc}>1$)~\cite{spenley2000}. Moreover, for the majority of nanoscale systems hydrodynamic flows occur in the low Reynolds number regime ($\tn{Re} < 1$) in which case inertia is irrelevant and thus an incorrect Schmidt number is not expected to affect any observables. 
If required, high Schmidt numbers can be obtained by increasing the friction coefficient~$\gamma$ or modifying the weighing function~$w(r_{ij})$, at a cost of efficiency~\cite{krafnick2015,lauriello2021}, or by using the Lowe thermostat~\cite{lowe1999}. We stress that the proposed DPDS method is general and can be applied to any DPD parameters that model higher Schmidt numbers or non-aqueous solutions.

The simulation time scale~$\tau$ is determined by setting the desired dynamic viscosity~$\eta$ of the fluid. The viscosity of the DPD fluid can be measured \textit{via} self-diffusivity, the decay of the stress autocorrelation function or the Poiseuille profile~\cite{boromand2015,lauriello2021}. At the standard parameters used the viscosity obtained by fitting the Poiseuille profile is $\eta=(2.31 \pm0.05) k_\tn{B}T \tau r_\tn{c}^{-3}$ (see Sec. Nanochannel flow), which is consistent with the value obtained by the decay of the stress autocorrelation function ($\nu_\tn{k}=0.748 k_\tn{B}T \tau m^{-1}$, at $\rho=3r_\tn{c}^{-3}$)~\cite{lauriello2021}. To model the dynamic viscosity of water, $\eta=10^{-3}\, \tn{Pa}\cdot\tn{s}$, at room temperature, $T=298$~K, the MD simulation time scale is thus $\tau=29\,$ps. 
~\footnote{Using this timescale we can determine the mass of the particles and find $m\approx8.3 10^{-24}\,$kg, which is about two orders of magnitude larger than the mass of three water molecules. This is a consequence of the low Schmidt number at the standard DPD parameters and the requirement to model the desired dynamic viscosity~$\eta$.}


\section{DPD solvent}
\label{sec:dpds}

We aim to incorporate hydrodynamic interactions within an existing system of solutes, such as ions, molecules, polymers, or nanoparticles.
The premise of the DPDS method that achieves this is:

(i) There are no conservative interactions between the system and the DPD fluid. This requirement ensures that all equilibrium configurational observables of the system are not affected by the presence of the DPD solvent. The DPD solvent parameters ($a_{ij}, \gamma$) are set to reproduce desired compressibility and viscosity of the solvent. 

(ii) The system interacts with the solvent \textit{via} the DPD thermostat whose coupling is set to reproduce the desired diffusion constant of the solutes. 
Specifically, a system containing $N_\tn{s}$ solute particles is described by interactions that define the potential energy $V_\tn{s}(\mathbf{r}^{N_\tn{s}}_\tn{s})$ depending on the positions $\mathbf{r}^{N_\tn{s}}_\tn{s}$ of these solute particles.
To preserve the equilibrium distribution of configurational states, the DPD fluid is coupled to the solutes only \textit{via} the random and dissipative forces [Eqs.~\eqref{eq:Fd} and~\eqref{eq:Fr}] and
the solute--DPD coupling is determined by the strength $\gamma_\tn{s}$ and the weighting function $w_\tn{s}(r_{ij})$. To simplify the implementation and reduce the number of parameters, $w_\tn{s}(r_{ij})$ has the same form as between DPD particles, Eq.~\eqref{eq:w}, but with a different cutoff value $r_\tn{s}$,
\begin{equation}
w_\tn{s}(r_{ij}) = 1-\frac{r_{ij}}{r_\tn{s}} \;,
\label{eq:ws}
\end{equation}
where the index $i$ refers to solutes and the index $j$ to solvent (DPD) particles. We stress that there is nothing special about this form, Eq.~\eqref{eq:ws} is chosen for convenience but any other peaked function could be used.
There are no direct dissipative or random forces between solvent particles.  
The perpendicular and parallel coupling strengths and weighing functions are chosen to be the same, $\gamma_{\perp,\tn{s}}=\gamma_\tn{s}$ and $w_{\perp,\tn{s}}(r_{ij})=w_\tn{s}(r_{ij})$.  Thus, the solute--solvent coupling is determined by just two parameters: strength~$\gamma_\tn{s}$ and range~$r_\tn{s}$. The range $r_\tn{s}$ is determined by the size of the solute including the solvation layer and $\gamma_\tn{s}$ by the desired solute diffusion constant~$D$. 

The solute--solvent hydrodynamic coupling [Eq.~\eqref{eq:ws}] is spherically symmetric, which is applicable to the large majority of coarse-grained models that are based on spherically symmetric excluded-volume interactions, such as monomers in a bead-spring polymer~\cite{stevens95}, ions, force-fields like MARTINI~\cite{marrink2007}, or DNA/RNA models such as oxDNA~\cite{snodin2015}. Non-spherically symmetric cases are discussed later.

 \begin{figure}
\centering
\includegraphics[width=\figurewidth]{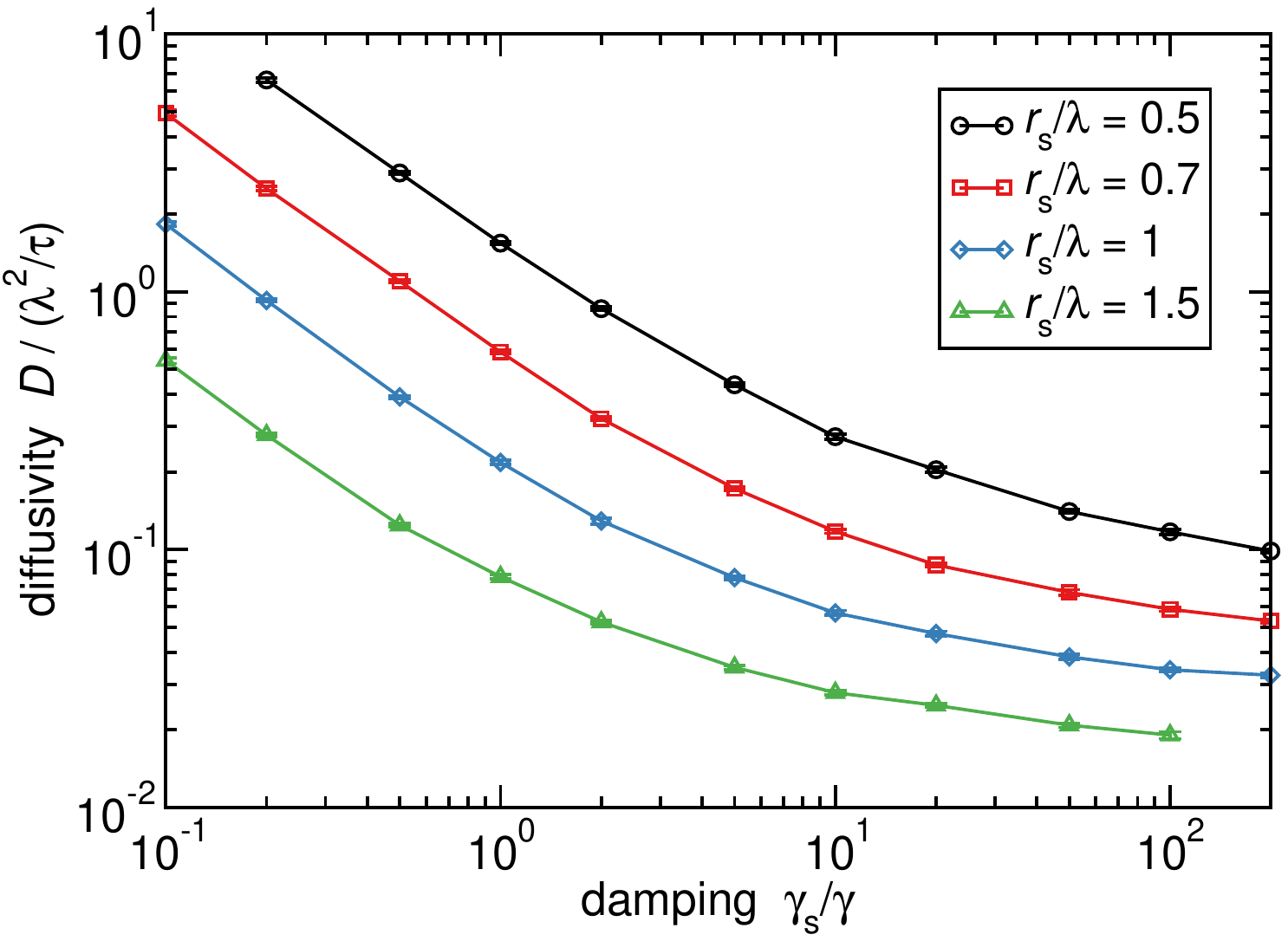}
\caption{Diffusion constant depending on the solute size $r_\tn{s}$ and coupling strength $\gamma_\tn{s}$. Data obtained by simulating system size $L=100\lambda$ at dilute solute density $\rho_\tn{s}=10^{-3}\lambda^{-3}$ and timestep $\Delta t=10^{-3}\tau$ at the standard DPD parameters: $r_\tn{c}=\lambda$, $\rho=3r_\tn{c}^{-3}$, $a_{ij}=78k_\tn{B}T$, $\gamma=4.5k_\tn{B}T\tau r_\tn{c}^{-2}$. $D=\langle r^2 \rangle/(6\tau)$ with the MSD $\langle r^2 \rangle$ calculated in the diffusive regime ($t>50\tau$).}
\label{fig:D}
\end{figure}

\begin{figure}
\centering
\includegraphics[width=\figurewidth]{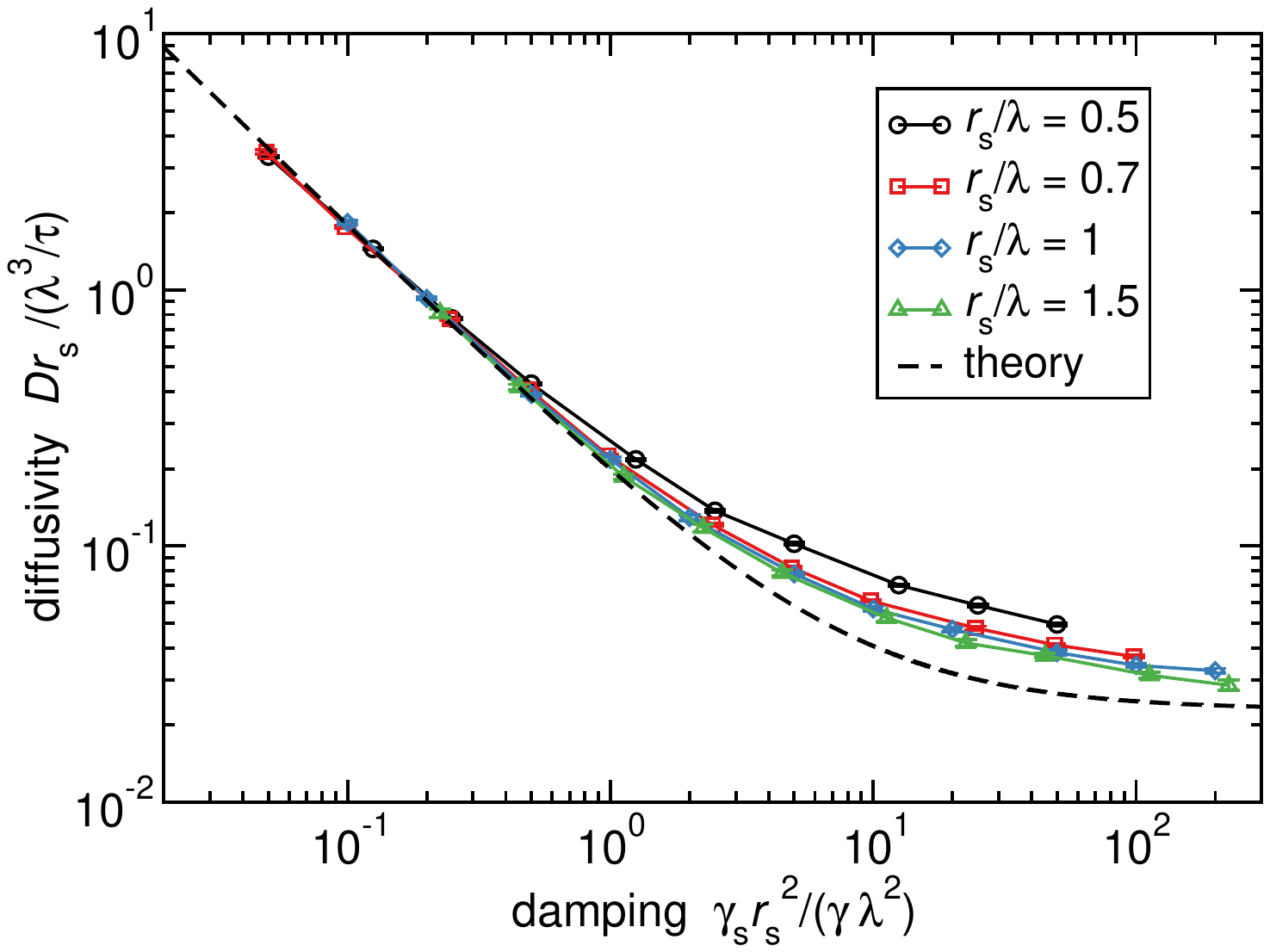}
\caption{Master plot of diffusivity data from simulations (symbols) and comparison to the analytical prediction (dashed line, Eq.~\eqref{eq:Dtheory}).}
\label{fig:Dmaster}
\end{figure}

The diffusion constant $D=D(r_\tn{s},\gamma_\tn{s})$ measured \textit{via} the mean-squared displacement is shown in Figure~\ref{fig:D}. 
The diffusion constant scaling can also be estimated analytically. Strong coupling, $\gamma_\tn{s}\to\infty$, implies the solvent is rigidly coupled to the solute within range $r_\tn{s}$, thus the solute behaves as a solid sphere with radius~$r_\tn{s}$ whose diffusivity is given by the Stokes-Einstein relation, $D=k_\tn{B}T/(6\pi \eta r_\tn{s})$. Conversely, in the weak coupling regime, the diffusion constant can be calculated by mapping the thermostat forces to a Langevin equation~\cite{groot1997}, which yields the diffusion constant $D=15k_\tn{B}T/2\pi\gamma_\tn{s}\rho r_\tn{s}^3$ (note that this expression is a factor 3 smaller than the result in Ref.~\cite{groot1997} due to inclusion of shear forces, $\gamma_{\perp,\tn{s}}=\gamma_\tn{s}$). These two limits imply the following functional form for the diffusion constant of the solutes,
\begin{equation}
D(r_\tn{s},\gamma_\tn{s})=\frac{k_\tn{B}T}{2\pi r_\tn{s}} \left(\frac{1}{3\eta} + \frac{15}{\rho \gamma_\tn{s} r_\tn{s}^2} \right) \;.
\label{eq:Dtheory}
\end{equation}
Using this scaling, the simulation data collapse to a master plot and follow Eq.~\eqref{eq:Dtheory} to within $\approx30\%$ (Fig.~\ref{fig:Dmaster}). Therefore Eq.~\eqref{eq:Dtheory} can be used to estimate $\gamma_\tn{s}$ given a desired diffusion constant~$D$ and solute size $r_\tn{s}$. If the thermostat without shear forces is used ($\gamma_{\perp,\tn{s}}=0$), the second term in Eq.~\eqref{eq:Dtheory} should be multiplied by a factor 3.
 
\begin{figure}
\centering
\includegraphics[width=\figurewidth]{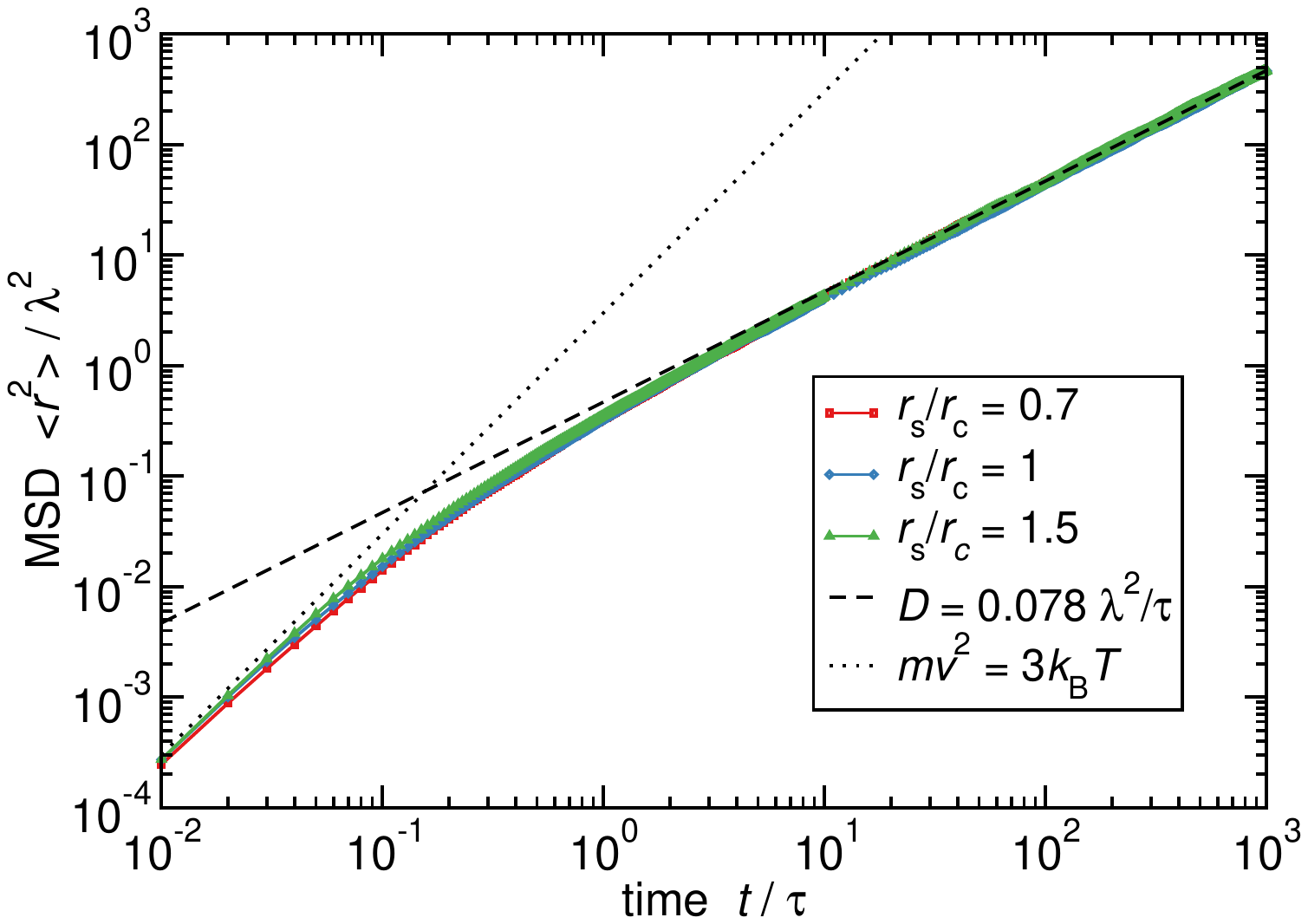}
\caption{Mean-squared displacement (MSD) of solutes in bulk solution for different $r_\tn{s}$ at $D\approx0.078\lambda^2/\tau$. [$r_\tn{s}$,  $\gamma_\tn{s}] = [0.7\lambda, 30\gamma]$ (red squares), $[\lambda, 5\gamma]$ (blue diamonds), $[1.5\lambda, \gamma]$ (green triangles) showing the ballistic-to-diffusive transition does not explicitly depend on $\gamma_\tn{s}$ and $r_\tn{s}$, but only on $D$.}
\label{fig:msd}
\end{figure}

The diffusion constant of a typical small solute such as an ion or a small molecule is $D\approx1~\tn{nm}^2/\tn{ns}$, which is $D\approx0.078 \lambda^2/\tau$ in the standard DPD units ($\lambda=0.646~\tn{nm}$, $\tau=0.029~\tn{ns}$). Our simulation data confirm that chosing the mass of the solute particles the same as the mass of the DPD particles the dynamics are already fully in the diffusive regime at $\langle r^2\rangle=\lambda^2$ (Fig.~\ref{fig:msd}).  As expected, the dynamics do not depend on specific $\gamma_\tn{s}$ and $r_\tn{s}$ as long as $D$ is kept constant. The transition from ballistic to diffusive dynamics occurs on the lengthscale $r_\tn{BD} = 2\sqrt{3}D\tau/\lambda \approx0.27\lambda$  (Fig.~\ref{fig:msd}).  Therefore, we conclude that the standard DPD friction $\gamma=4.5k_\tn{B}T\tau r_\tn{c}^{-2}$ models the appropriate diffusive dynamics on the relevant lengthscales ($r \ge \lambda$). 

If necessary, the ballistic--diffusive transition $\tau_\tn{BD}$ can be pushed to even smaller values by increasing the DPD solvent friction~$\gamma$ or decreasing the solute mass, but both would require a smaller integration time-step decreasing the efficiency. Conversely, if a larger ballistic regime is permitted, $\gamma$ could be decreased which would reduce the DPD viscosity and thus increase the value of the time unit~$\tau$, allowing exploration of longer timescales, but that would also reduce the Schmidt number (Sc), which may be important, see discussion below.

\subsection{Limitation of the method: solute permeability}

 \begin{figure}
\centering
\includegraphics[width=\figurewidth]{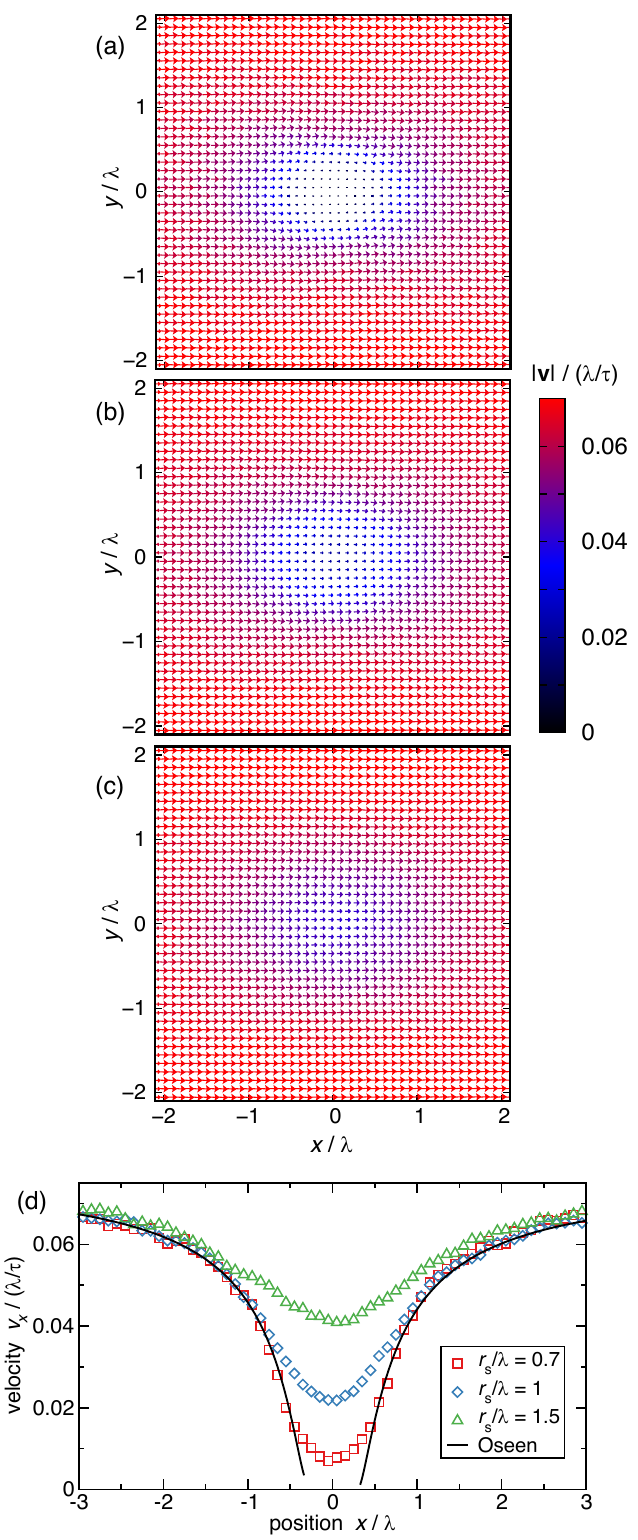}
\caption{Velocity profile around a spherical solute located at $x=y=z=0$. (a,b,c) The three profiles are obtained at [$r_\tn{s}$,  $\gamma_\tn{s}] = [0.7\lambda, 30\gamma]$ (a), $[\lambda, 5\gamma]$ (b), $[1.5\lambda, \gamma]$ (c) corresponding to $D=0.078\lambda^2/\tau$ (Fig.~\ref{fig:msd}). The profiles are calculated by averaging the velocity of all DPD particles within $-0.1<z/\lambda<0.1$ for $t=10^7\tau$. System size $L=10\lambda$ with periodic boundary conditions and body force density $G_x=10^{-3}k_\tn{B}T/\lambda^4$ imposed on DPD particles. (d) The velocity $v_x$ on the center line is obtained by averaging within $-0.1<y/\lambda<0.1$ and $-0.1<z/\lambda<0.1$.  
The Oseen analytical prediction (black line) is given by Eq.~\eqref{eq:Oseen}. 
}
\label{fig:stokes}
\end{figure}

The main limitation of using only the dissipative and random forces for the solute--solvent interaction is that solutes are not impermeable to the solvent particles. The coupling~$\gamma_\tn{s}$ effectively determines the local viscosity at the location of the solute particle, but the solvent can pass through the solute.  This is illustrated by the velocity profiles around a spherical solute (Fig.~\ref{fig:stokes}). Larger coupling $\gamma_\tn{s}$ leads to smaller fluid velocity at the particle location. In the limit $\gamma_\tn{s}\to\infty$ the solute becomes effectively impermeable. In this sense, DPDS bears similarity to the MPC~\cite{gompper2009} and FPD~\cite{tanaka2000} methods.

The effect of permeability can be systematically investigated by choosing different $r_\tn{s}, \gamma_\tn{s}$ that yield the desired diffusion constant (Fig.~\ref{fig:stokes}).  At large $\gamma_\tn{s}$ the flow profile approaches the Oseen prediction for an impermeable spherical particle~\cite{batchelor1967},
\begin{eqnarray}
v_x^\tn{b} &=& v_0 \left( 1+ \frac{a^3}{2x^3}- \frac{3a}{2x} \right) \nonumber , \\
v_x^\tn{f} &=& v_0 \left( 1+ \frac{a^3}{2x^3} - \frac{3a^2}{2x^2\tn{Re}}\left[1-e^{-x\tn{Re}/a}\right] \right) ,
\label{eq:Oseen}
\end{eqnarray}
where $v_x^\tn{b}$ and $v_x^\tn{f}$ are the flow velocities behind and in front of the particle on the symmetry axis at $z=y=0$. $v_0=D f_\tn{particle} /[k_\tn{B}T(1+3\tn{Re/8})]$ is the far field velocity at total force $f_\tn{particle}=G_x L^3$ with $G_x$ the body force on the fluid and $L$ the system size (to calculate the flow profiles the particle is immobilized and the force is applied on the fluid).  The radius of a sphere that yields the desired diffusion constant $D$ is $a=k_\tn{B}T/(6\pi\eta D)$ and the corresponding Reynolds number $\tn{Re}=a\rho v_0/\eta$. 
The simulated flow profiles approach the Oseen prediction for large $\gamma_\tn{s}$ (Fig.~\ref{fig:stokes}d). 
The ratio $\gamma_\tn{s}/\gamma=5$ appears to be sufficiently large to reproduce the Oseen profile to a distance $r\approx  3/4\lambda$ and also the hydrodynamic scaling of polymer collapse as shown below. 

Another limitation of solute permeability concerns fast time-dependent changes to the flow. The timescale of the solute--solvent coupling was calculated above, $\tau_\tn{BD} = 2 D \tau^2/\lambda^2$, and thus any temporal momentum change on timescales $t \le \tau_\tn{BD}$ will not be adequately transferred between solute and solvent. However, for standard DPD parameters and typical solute diffusivities we find $\tau_\tn{BD} < \tau $ (Fig.~\ref{fig:msd}); the timescale of solute--solvent coupling is smaller than that of solvent--solvent coupling. Therefore, DPDS is not expected to introduce additional high-frequency limitations and the standard DPDS parameters used here are sufficient for any perturbations slower than $\tau_\tn{BD}\approx10~\tn{ps}$, which should be adequate for most applications. If even faster response times are required, $\tau$ can be reduced by increasing the thermostat friction $\gamma$.

If permeability is not allowed, for example to model impermeable membranes, the repulsive interactions between solute and solvent can be added to the model. This is partially achieved by the standard DPD approach where solute--solvent interaction, e.g. for membrane lipids~\cite{groot2001} or polymers~\cite{Guo2011}, is also represented as soft repulsion [Eq.~\eqref{eq:V}]. Full blocking could be achieved by hard repulsive interaction such as Lennard-Jones or WCA. In either case, the solute--solvent conservative interactions change the equilibrium distributions of the solutes and induce entropic depletion interactions between solutes if solutes and solvent particles are not of the same size. This likely requires that the solute model is designed and parametrized for specific DPD solvent parameters.

\subsection{Recipe}

We provide a straightforward recipe on how to use the DPDS method with coarse-grained models of solutes.
\begin{enumerate}
\item Choose the desired solvent properties: determine the minimum DPD length scale~$r_\tn{c}$ on which to resolve hydrodynamics and obtain the desired compressibility \textit{via} $a_{ij}$~\cite{groot1997}. Chose the thermostat coupling~$\gamma$ that results in the desired transition between ballistic and diffusive regimes. The standard parameters, $r_\tn{c}=0.646~$nm, $\rho=3r_\tn{c}^{-3}$, $a_{ij}=78k_\tn{B}T$,  $\gamma=4.5 \sqrt{k_\tn{B}T m}/r_\tn{c}$ are likely a good starting point for most coarse-grained models of aqueous solutions of molecules, ions, and polymers.

\item Determine the solute--solvent coupling, $r_\tn{s}$ and $\gamma_\tn{s}$ that yield the desired solute diffusivity~$D$: The size $r_\tn{s}$ should be similar to the physical size of the solutes and $\gamma_\tn{s}$ is determined \textit{via} Eq.~\eqref{eq:Dtheory} or by measuring the mean-squared displacement of solutes. 

\end{enumerate}

This introduces both hydrodynamics interactions and thermostating of the solute system, while maintaining the equilibrium configurational distribution of the solutes.
An example implementation in the open-source MD package LAMMPS is provided in the Appendix.

\section{Applications}
\label{sec:appl}

 To demonstrate the applicability of the DPDS method, we investigate two different systems where hydrodynamic interactions play a crucial role: the collapse and diffusion dynamics of a single polymer and electroosmotic flow of an electrolyte solution. 

\subsection{Polymer dynamics}

We consider a bead--spring polymer model~\cite{stevens95} in an aqueous solution. Consecutive beads in a polymer chain of $N$ beads are connected \textit{via} a harmonic potential
\begin{equation}
U=\frac{K}{2}(r_{ij}-\lambda)^2 
\end{equation}
with zero-energy bond length~$\lambda$ and strength $K=100k_\tn{B}T/\lambda^2$. 

The polymer is immersed in a DPD solvent  described by the standard parameters for an aqueous solution ($\rho=3r_\tn{c}^{-3}$, $\gamma=4.5k_\tn{B}T \tau/r_\tn{c}^2$, $a_{ij}=78k_\tn{B}T$, $r_\tn{c}=\lambda=0.646$~nm). The coupling between the polymer beads and the DPD fluid is achieved with $\gamma_\tn{s}=5\gamma$ and $r_\tn{s}=r_\tn{c}$, which models the diffusion constant of individual monomers $D=1.1~\tn{nm}^2/\tn{ns}$~(Fig.~\ref{fig:D}). The system is evolved using the velocity-Verlet integrator with time step $\Delta t=0.005\tau$.

First, we measure the diffusion constant for different polymer sizes and compare the scaling with the Zimm dynamics prediction $D \propto N^{-\nu}$, with $\nu$ the scaling exponent; $\nu=0.588$ at good-solvent conditions~\cite{RubinsteinColbyBook}.
The system is a cubic box with size $L=100\lambda$ and periodic boundary conditions. Polymers of length $N$ and monomer density $\rho_\tn{m}=10^{-3}/\lambda^3$ are equilibrated at good solvent conditions. Bead--bead repulsion is modeled as  a Lennard-Jones (WCA) interaction with size $\sigma=\lambda$, strength $\epsilon={k_\tn{B}T}$ and cutoff $r_\tn{cut}=2^{1/6}\sigma$. We measure the mean-squared-displacement of all monomers~$\langle r^2 \rangle$ for $t=5000\tau$. The diffusion constant is measured \textit{via} the slope of the average mean squared-displacement as a function of time, $6D = \partial \langle r^2 \rangle/\partial t$; at large displacements, $\langle r^2 \rangle > (2R_\tn{g})^2$.

\begin{figure}
\centering
\includegraphics[width=\figurewidth]{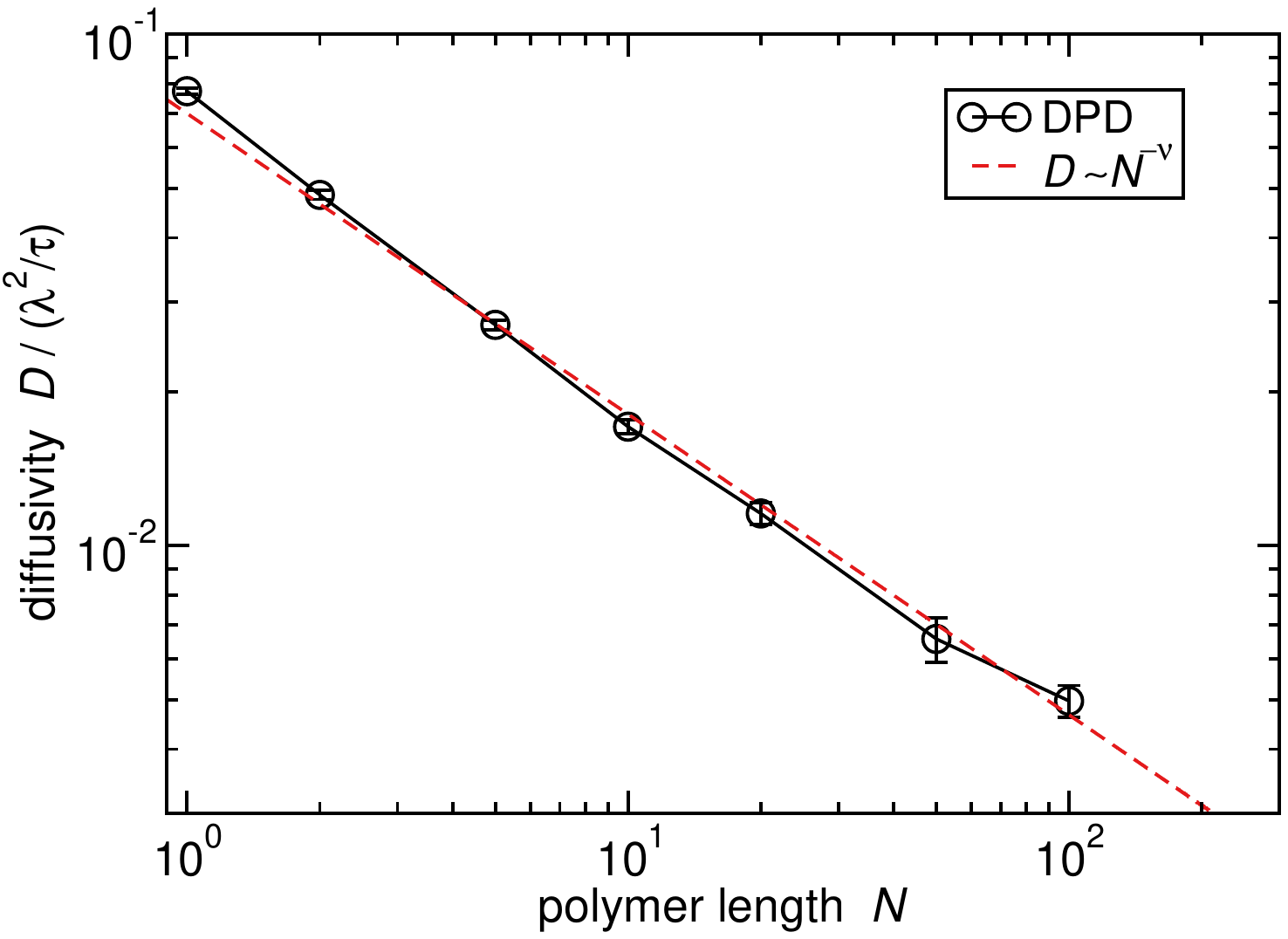}
\caption{Diffusion constant~$D$ dependence on polymer size~$N$. Simulation data (black circles) is well described by Zimm dynamics (red dashed line) with the good-solvent scaling exponent $\nu=0.588$. Error bars denote standard errors obtained from 8 independent simulations.}
\label{fig:zimm}
\end{figure}

The scaling of the diffusion constant with size fully reproduces the expected Zimm dynamics (Fig.~\ref{fig:zimm}). Small deviations ($<10\%$) occur for very short polymers ($N\sim 1$) due to finite size effects.

We next measure the timescale of a hydrophobic collapse of the polymer, a problem related to the dynamics of protein folding.  Analytical predictions for the hydrophobic collapse timescale~$\tau_\tn{c}$ of an initially expanded polymer indicate $\tau_\tn{c}\sim N^2$ for brownian dynamics and $\tau_\tn{c} \sim N^{4/3}$ with hydrodynamics~\cite{Kikuchi2005}, since a polymer globule needs to travel a distance $\sim N$ while the drag on the collapsed globules scales as $\sim N$ for brownian dynamics and $\sim N^{1/3}$ for Stokes flow. 

The collapse timescale~$\tau_\tn{c}$ is defined by the time required for the change in the radius of gyration~$R_\tn{g}(t)$ to reach a fraction $f_\tn{c}=0.9$ of the maximum change,
\begin{equation}
R_\tn{g}(\tau_\tn{c})=(1-f_\tn{c})R_\tn{g,0} + f_\tn{c} R_\tn{g,col} \;,
\label{eq:Rgcol}
\end{equation}
where $R_\tn{g,0}$ and $R_\tn{g,col}$ are, respectively, the initial and the final (collapsed) $R_\tn{g}$ values. 
We consider two initial configurations, (i) a fully expanded linear polymer ($R_{\tn{g},0}\sim N$) and (ii) a polymer equilibrated at good solvent conditions ($R_{\tn{g},0}\sim N^{\nu}$) with the scaling exponent $\nu=0.588$. 
To model the collapse, we introduce attractive interactions between monomers modeled by the Lennard-Jones potential with size $\sigma=\lambda$, strength $\epsilon=1.25k_\tn{B}T$, and cutoff $r_\tn{LJcut}=2.5\sigma$, which introduces a sudden quench in solvent quality. We can estimate the Flory-Huggins $\chi$ parameter by matching the critical point of the Lennard-Jones fluid, $T_\tn{c,LJ} \approx 1.3 \epsilon/k_\tn{B}$, to the regular solution model, yielding $\chi \approx 2 T_\tn{c,LJ}/T$. For $\epsilon=1.25k_\tn{B}T$, this results in $\chi\approx3.25$, which is sufficiently large to drive polymer collapse. 

\begin{figure}
\centering
\includegraphics[width=\figurewidth]{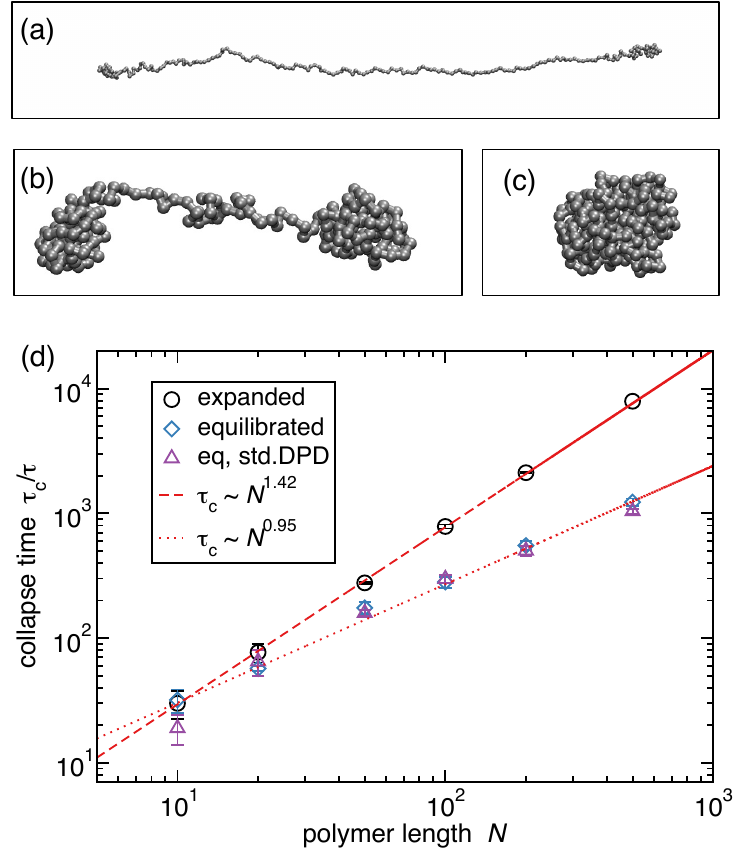}
\caption{Polymer collapse. (a,b,c) configurations of a bead-spring polymer of size $N=200$ during collapse from an initially fully-expanded linear chain at (a) $t=500\tau$, (b) $t=2000\tau$, and (c) $t=2500\tau$. (d) DPDS simulation data showing collapse timescale dependence on the polymer size~$N$ for an initially expanded linear chain (black circles), which follows $\sim N^{1.42\pm0.02}$ scaling, and an initially equilibrated polymer at good solvent conditions (blue diamonds), which follows $\sim N^{0.95\pm0.05}$. For comparison, we show the collapse data using the standard DPD model with soft-repulsive beads (purple triangles). System size for initially expanded case, $L_x=N+10\lambda$, $L_y=L_z=50\lambda$, and for equilibrated polymers $L_x=L_y=L_z=80\lambda$. Error bars denote standard errors obtained from 5 (black circles) and 20 (blue diamonds and purple triangles) independent simulations.}
\label{fig:collapse}
\end{figure}

Simulation data shown in Fig.~\ref{fig:collapse} show the scaling follows $\tau_\tn{c} \sim N^{1.42\pm0.02}$ for an initially fully stretched polymer, while $\tau_\tn{c} \sim N^{0.95\pm0.05}$ for an initially equilibrated polymer. The first scaling exponent is close to the analytical prediction for an expanded polymer $N^{4/3}$, and the second agrees with previous DPD predictions for an equilibrated polymer~\cite{Guo2011},  $\tau_\tn{c} \sim N^{0.98\pm0.09}$.

To further validate the DPDS method we compare it with the standard DPD approach. We perform collapse simulations of a neutral polymer using the standard DPD model where the conservative monomer--monomer and monomer--solute interactions are described by the DPD potentials [Eq.~\eqref{eq:V}] and thermostat with $\gamma=4.5k_\tn{B}T\tau r_\tn{c}^{-2}$ is applied between all pairs of particles. The interaction strength is described by the prefactor $a_{ij}$, where $i$ denotes a solute and $j$ a solvent. We use $a_{jj} = a_{ij} = 78k_\tn{B}T$. The polymer is equilibrated in good solvent conditions ($a_{ii}=85 k_\tn{B}T$) and the collapse is then initiated by changing the monomer--monomer interaction to $a_{ii}=65 k_\tn{B}T$, thereby reducing the solvent quality. The interaction difference $a_{ij}-a_{ii}=13k_\tn{B}T$ resulting in the $\chi$ parameter~\cite{groot1997} of $\chi\approx3.5$ that is very close to above estimate for the Lennard-Jones model. The collapse timescales using the standard DPD approach agree with the DPDS method and Lennard-Jones polymer model (Fig.~\ref{fig:collapse}).

This data on polymer diffusion and collapse timescale indicates that the DPDS method faithfully reproduces the hydrodynamic coupling between a polymer and a solvent. Contrary to standard DPD, the DPDS method combined with Lennard-Jones interaction (or a similar hard repulsive interaction) allows straightforward addition of point charges and simulation of polyelectrolytes~\cite{Yuan2024}.

\subsection{Nanochannel flow}

We next consider the electro-osmotic flow of an electrolyte solution in a slit nanochannel and investigate the coupling between hydrodynamics and electrostatic interactions. To use DPDS for wall-bounded flows, we must first briefly discuss a method to impose a desired no-slip or slip boundary condition at the channel walls.

\subsubsection{wall boundary condition}

To impose the desired boundary condition at the channel walls, we investigate a pure solvent system without ions. Implementation of solid walls within DPD simulations is not straightforward due to layering artifacts that can occur next to a flat wall~\cite{pivkin2005}. A no-slip boundary condition can be imposed by introducing a layer of immobilized DPD particles at the walls~\cite{bercelos2021}, however, due to repulsive interactions, the slip length dependence on the wall particle density is non-monotonic. Another possibility is to impose a drag force parallel to the wall~\cite{smiatek2008,smiatek2009}. 

Here we propose an alternative strategy to impose a boundary condition by coupling the DPD fluid to the immoblized wall particles only through the thermostat with coupling strength~$\gamma_\tn{w}$, which is determined by the desired slip-length. The method is very similar to using a parallel drag force~\cite{smiatek2008}, but is expected to be easier to employ because it does not require a separate implantation of a wall--DPD thermostat.

 The DPD solvent particles interact with a smooth repulsive surface \textit{via} a repulsive WCA interaction with $\sigma=r_\tn{c}$ and $\epsilon=k_{\tn B}T$. The smooth repulsive surfaces are positioned at $y_\tn{wall} = \pm (w/2+r_\tn{c})$, where $w$ is the width of the nanochannel. In addition, immobile particles are placed at the wall surface~$y_{\tn{im}}=\pm w/2$, at 2D density $\rho_{\tn w}=\rho r_\tn{c}$ arranged on a regular mesh with lattice spacing $a_\tn{w}=\rho_{\tn w}^{-1/2}$. These immobile particles interact with the DPD particles only \textit{via} the DPD thermostat with strength $\gamma_\tn{w}$ and range $r_\tn{c}$ [Eqs.~\ref{eq:Fd}, \ref{eq:Fr}, \ref{eq:w}]. 
 
\begin{figure}[h]
\centering
\includegraphics[width=\figurewidth]{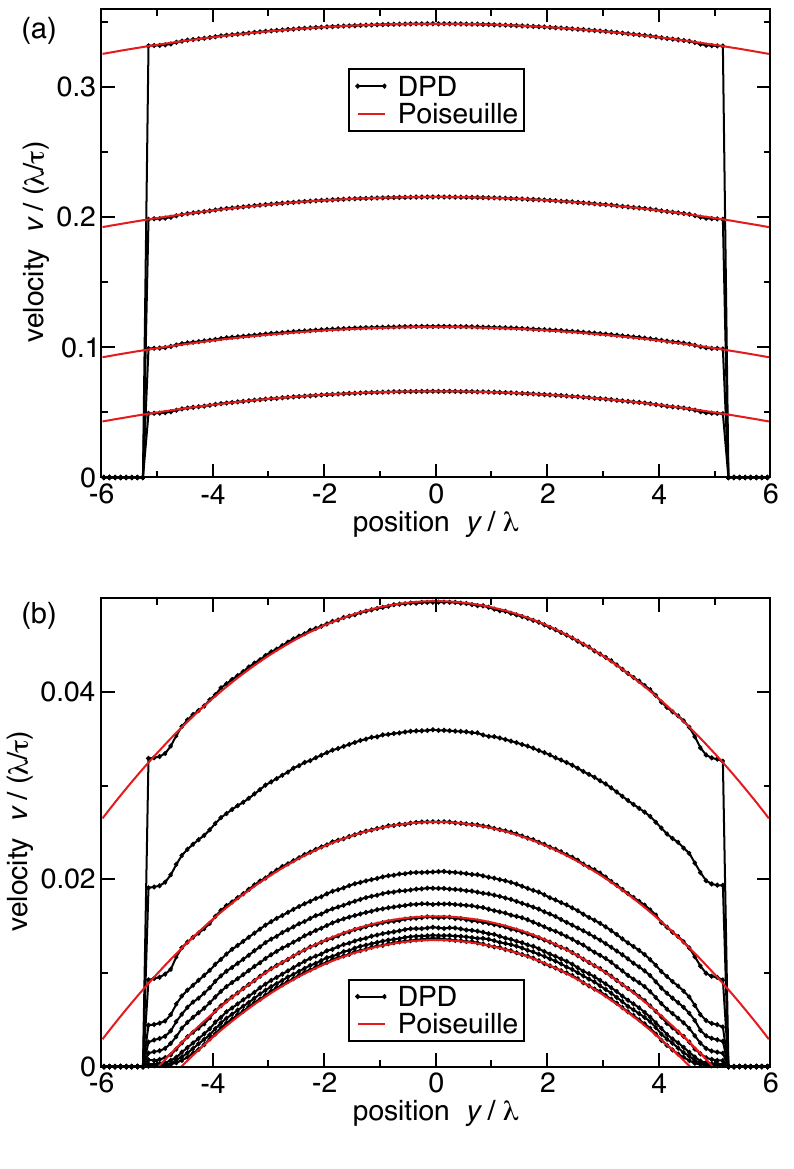}
\caption{Velocity flow profile from DPD simulations (black circles) and comparison to Poiseuille profiles (red curves) given by Eq.~\eqref{eq:Pois-slip2}. (a,b) DPD data (black) from bottom to top are obtained at $\gamma_\tn{w}/\gamma=[5,3,2,1,0.5,0.3,0.2,0.1,0.05,0.03,0.02,0.01,0.005,0.003]$ at pressure gradient $G_x=0.003k_\tn{B}T/\lambda$ and $w=10\lambda$. Poiseuille prediction (red curves) from bottom to top shown for slip length $L_\tn{s}/\lambda=[-0.42,-0.03,1.52,5.15,7.7,15.3,30.7,51.2]$ at viscosity $\eta=2.31 k_\tn{B}T\tau/\lambda^3$ [Eq.~\eqref{eq:Pois-slip2}]. }
\label{fig:pois}
\end{figure}

 To induce flow, a pressure gradient~$G_x$ is imposed on the fluid as a body force, $f_x=G_x/\rho$, which acts on each DPD particle. The system size is $L_y=L_z=w$ and $L_x=40 \lambda$ with periodic boundary conditions in $x$ and $z$. The system is simulated for $t_\tn{init}=5000\tau$ to reach steady state followed by $t=5\cdot10^5\tau$ for calculation of the velocity profiles at different values of the wall--solvent coupling~$\gamma_\tn{w}$ (Fig.~\ref{fig:pois}).

The velocity profile between two parallel plates at low Reynolds numbers follows the parabolic Poiseuile profile,
\begin{equation}
v(y) = \frac{G_x}{2\eta} (w^2/4 - y^2 + w L_{\tn s}) .
\label{eq:Pois-slip2}
\end{equation}
The slip length
\begin{equation}
L_\tn{s}=\pm{\Delta v} \left[\left(\frac{\partial v}{\partial y}\right)_{y=\mp w/2}\right]^{-1} \;,
\label{eq:Ls}
\end{equation}
is determined by the slip velocity $\Delta v$ at the wall. We find the simulated velocity profiles accurately reproduce the parabolic Poiseuille profiles for at least two orders of magnitude in flow velocity and slip lengths from zero to larger than channel width (Fig.~\ref{fig:pois}). 
A small deviation from the parabolic profile is observed within a distance $r_\tn{c}$ of the wall due to the layering effect of the DPD particles next to a smooth repulsive wall.

\begin{figure}
\centering
\includegraphics[width=\figurewidth]{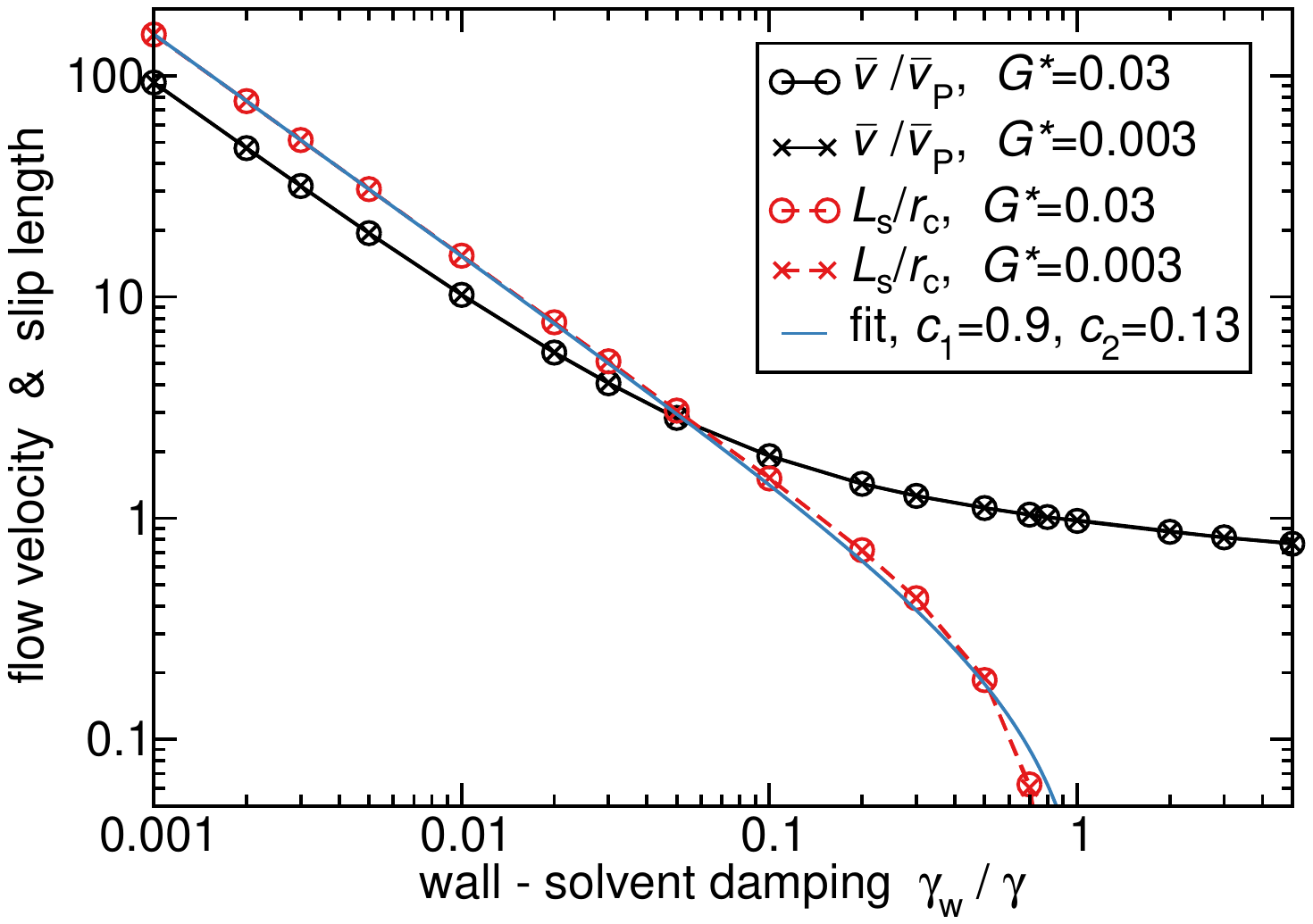}
\caption{Average velocity obtained from simulations (solid black line) and corresponding slip length using Eq.~\eqref{eq:Ls2} (dashed red lines) using viscosity $\eta=2.31 k_\tn{B}T\tau/\lambda^3$. The data for different pressure gradients $G^*=G_x/k_\tn{B}T\rho\lambda$ show perfect overlap. The two parameter fit [Eq.~\eqref{eq:Lstheory}](solid blue line) is accurate to within $0.1r_\tn{c}$. }
\label{fig:dpd-slip2}
\end{figure}

To avoid numerical errors when calculating derivatives [Eq.~\eqref{eq:Ls}], we determine the slip length from averages in the velocity. Since the velocity profiles are parabolic (Fig.~\ref{fig:pois}), the slip length can be obtained by integrating the profile [Eq.~\eqref{eq:Pois-slip2}],
\begin{equation}
L_\tn{s} = \left(\frac{\bar{v}}{\bar{v}_\tn{P}}-1\right) \frac{w}{6}
\label{eq:Ls2}
\end{equation}
where $\bar{v}$ is the average velocity in the channel and 
\begin{equation}
\bar{v}_\tn{P}=\frac{G_xw^2}{12\mu}
\label{eq:barvp}
\end{equation}
 is the average velocity for Poiseuille profile at $L_\tn{s}=0$. Thus, we obtain the slip length as a function of wall damping~$\gamma_{\tn w}$ (Fig.~\ref{fig:dpd-slip2}). 
 
 Analytical considerations show that to second order the slip length is given by
 \begin{equation}
 L_\tn{s}/r_\tn{c} = c_1/\alpha - c_2 - \mathcal{O}(\alpha)
 \label{eq:Lstheory}
 \end{equation}
 with $\alpha=r_\tn{c}^2\gamma_\tn{w}\rho/\eta$~\cite{smiatek2008} and the positive constants $c_1, c_2$ of order unity that depend on the wall solvent interaction. We find that fitting $c_1=0.9$ and $c_2=0.13$ can be used to predict $L_\tn{s}$ to accuracy within $0.1r_\tn{c}$  (Fig.~\ref{fig:dpd-slip2}). 

\subsubsection{electroosmotic flow}

Having described the channel setup and the wall interaction, we show how to simulate hydrodynamic flow of an electrolyte. Free monovalent ions are modeled as charged spheres with the short-range ion--ion repulsion modeled by the WCA potential with hydrated ion diameter $\sigma=\lambda=0.646~\tn{nm}$, and interaction strength $\varepsilon=k_{\tn B}T$. The experimentally measured value for small-ion diffusivity, $D\approx \tn{nm}^2/\tn{ns}$, is obtained at solvent--solute coupling strength $\gamma_\tn{s}=5\gamma$ and range $r_\tn{s}=\lambda$ (Fig.~\ref{fig:msd}). The electrostatic interactions are calculated using PPPM Ewald summation with real-space cutoff $r_{\tn{ewald}}=5\lambda$ and relative force accuracy of $10^{-4}$. Slab correction factor 3.0 is used in the $y$ coordinate to simulate fixed boundary conditions at the channel walls. Electrostatic strength is determined by the Bjerrum length $l_\tn{B}=0.71\,$nm corresponding to an aqueous solution at room temperature.

\begin{figure}
\centering
\includegraphics[width=\figurewidth]{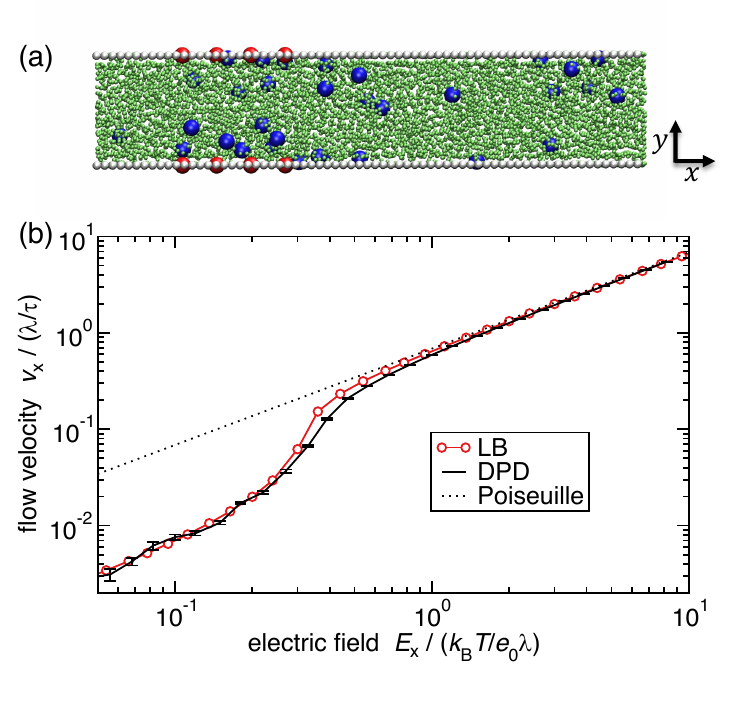}
\caption{electroosmotic flow in a slit channel. (a) Configuration setup showing DPD particles (green), immobile wall particles (white), surface charge (red), and counterions in (blue). (b) Mean velocity in a slit channel obtained from DPD simulations and comparison to lattice-Boltzmann (LB) and Poiseuille prediction [Eq.~\eqref{eq:barvp}] at driving force $G_x=E_x e_0 \rho_\tn{ion}$. Slip length $L_\tn{s}\approx30\lambda$ is obtained at $\gamma_\tn{w}=0.005\gamma$ in DPD, and fractional bounce-back $f_\tn{bb}=0.99$ in LB.}
\label{fig:eos}
\end{figure}

We consider an electroosmotic flow in a nanochannel of width $w=8\lambda$ under an external electric field $E_x$. To investigate a non-symmetric case where accurate description of convection and diffusion of ions is important, the wall contains a small charged section that covers a fraction $f_\tn{w}=0.25$ of the wall surface with charge density $\sigma_\tn{q}=0.2 e_0/\lambda^2$ (see Fig.~\ref{fig:eos}a). The solution contains counter-ions at density $\rho_\tn{ion}=2\sigma_\tn{q}f_\tn{w}/w$. This configuration introduces a pattern in the charge density and Poisson-Boltzmann (PB) calculations predict two distinct regimes due to the localization of counter-ions at small external fields~\cite{curk2024}. This effect is pronounced at non-zero slip lengths and we use $L_\tn{s}=30\lambda$, a typical order of magnitude for a slip length of an aqueous solution~\cite{joseph2005}. 

The DPDS results are in an excellent agreement with PB simulations and clearly show two distinct flow regimes (Fig.~\ref{fig:eos}). The PB calculations are performed by employing the \textit{Ludwig} open-source package with electrokinetics~\cite{capuani2004} using lattice size $\Delta x=\lambda/2$, reduced viscosity $\eta^*=0.2$ and parameters corresponding to kinematic viscosity of water $\nu_\tn{k}=10^{-6}~\tn{m}^2/\tn{s}$. The desired slip length is obtained \textit{via} a fractional bounce-back boundary condition~\cite{wolff2012} with fraction $f_\tn{bb}$. All other parameters are the same as described for DPD. Small deviations between DPD and PB occur at the transition between the two flow regimes, which we attribute to the lack of thermal fluctuations in the PB model, as well as the lattice approximation and the associated lack of unit charge discretization in PB. However, the flow velocity in the two regimes is in perfect agreement.

These results demonstrate the applicability of DPDS to model hydrodynamics of electrolyte solutions. When considering highly concentrated solution, the viscosity typically increases with electrolyte concentration. This effect is captured by the DPDS method, and the rate of viscosity increase with concentration depends on the solute--solvent thermostat coupling length $r_\tn{s}$ [Eq.~\eqref{eq:ws}]. Thus, the value of $r_\tn{s}$  could be chosen to obtain the desired quantitative relation between viscosity and solute concentration. Moreover, direct thermostat coupling [Eqs. \eqref{eq:Fd} and~\eqref{eq:Fr}] between solutes could be added to further increase the viscosity of highly concentrated solutions. 

\section{summary}

In summary, we proposed a DPD-solvent (DPDS) method that introduces solvent hydrodynamics to coarse-grained models of solutes. 
The solute--solvent interaction occurs only \textit{via} the dissipative and random forces, which ensures the equilibrium configurational properties of the solute system are not affected by the presence of the DPD solvent. The solute--solvent coupling strength is determined by the desired diffusion constants of the solute. Because the method is based on short-ranged DPD interactions, the computational cost scales linearly with the system size~\cite{frenkel-smit2}.

The DPDS method can be utilized as a replacement for a Nos\'e-Hoover or Langevin thermostat in coarse-grained MD simulations, while capturing the hydrodynamic interactions at desired solvent compressibility, viscosity and solute diffusivity (see \emph{Recipe} section). The examples shown demonstrate the method reproduces the correct hydrodynamic of Oseen flow, channel flow, electroosmotic effects, and polymer hydrodynamics. Moreover, since the method utilizes the standard DPD thermostat, the simulations can be performed using existing implementations in open-source molecular dynamics packages such as LAMMPS and ESPResSo. Thus, the method should be broadly useful as means to introduce hydrodynamics to existing coarse-grained models of molecules and soft materials.

The method could be used with multicomponent solvents that are described by two or more different types of DPD particles~\cite{pagonabarraga2001,merabia2008}. In this case, the solute--solvent coupling could be distinct among different solvents, modeling different diffusivities. In addition, the chemical potential difference of the solutes in different solvents could be introduced by soft solute--solvent interactions. 
While we have only considered solute models based on spherical excluded-volume interactions, such as ions or monomers in a polymers, the method can be applied to non-spherically symmetric solutes, for example, by uniformly distributing ghost particles inside a non-spherical object and coupling these ghost particles to the DPD solvent \textit{via} Eqs.~\eqref{eq:Fd} and \eqref{eq:Fr}.

\begin{acknowledgments}
I thank Ignacio Pagonabarraga, James D. Farrell and Jiaxing Yuan for discussions on comments on the manuscript. This work was supported by the startup funds provided by the Whiting School of Engineering at JHU and performed using the Advanced Research Computing at Hopkins (rockfish.jhu.edu), which is supported by the National Science Foundation (NSF) grant number OAC 1920103. 
 \end{acknowledgments}

\appendix*
\section{A}
Implementation of the DPDS in the LAMMPS open-source MD package can be achieved using the existing \texttt{dpd} or \texttt{dpd/ext} pair styles~\cite{lammps2022}. Let us assume that type 1 is the solvent and type 2 is the solute and hybrid/overlay is used to combine DPD interactions with the solute--solute interactions. For the solute--solvent, polymer--solvent and ion--solvent interaction cases discussed in Sec.~\ref{sec:dpds} and~\ref{sec:appl} (using the LJ reduced units with unit length $\lambda=r_\tn{c}=1$ and DPD particle density $\rho=3 \lambda^{-3}$), the solvent properties are determined by the DPD model:
\\
\\
 \texttt{pair\_coeff   1 1    dpd/ext~$a_{ij}$~$\gamma$~$\gamma_{\perp}$~1~1~1.0}
 \\
 \\
 while the solute--solvent coupling is determined by the thermostat:
 \\
 \\
 \texttt{pair\_coeff   1 2    dpd/ext/tstat~$\gamma_\tn{s}$~$\gamma_{\perp,\tn{s}}$~1~1~$r_\tn{s}/\lambda$} 
\\
\\
Combining these instructions with the desired solute--solute interactions and a velocity-Verlet integrator introduces hydrodynamic interactions without affecting the equilibrium canonical distribution of the solute particles as compared to using a Langevin or Nos\'e-Hoover thermostat. 


%

\end{document}